\documentclass[NewJournal,InsideFigs,SingleSpace,11pt,NoLists,NoLineNumbers]{ascelike_rev}
\usepackage{textcomp}
\usepackage{amsmath}
\usepackage{amssymb}
\usepackage{graphicx}
\usepackage{esint}

\begin{document}
\title{Investigation of Cyclic Liquefaction with Discrete Element Simulations}
\author{
Matthew R. Kuhn%
%
\thanks{%
Professor, Dept.\ of Civil Engrg.,
Donald P.\ Shiley School of Engrg., Univ.\ of Portland,
5000 N.\ Willamette Blvd.,
Portland, OR  97203 (corresponding author). E-mail: kuhn@up.edu.},
\ M.~ASCE;
\\
Hannah D. Renken%
\thanks{%
Civil Engineer, Federal Aviation Administration, Seattle, WA. 
Formerly research student, Univ. of Portland.};
\\
Austin D. Mixsell%
\thanks{%
Civil Engineer, Federal Aviation Administration, WSA, NAVAIDS Engineering
Center, Seattle, WA. 
Formerly research student, Univ.\ of Portland.}; and
\\
Steven L. Kramer%
\thanks{%
  John R.\ Kiely Professor of Civil and Environmental Engineering, 
  Dept.\ of Civil and Environmental Engrg., Univ.\ of
  Washington, Seattle, WA 98195. E-mail: kramer@u.washington.edu.
  },
\ M.~ASCE
}
\maketitle
\begin{abstract}
A discrete element (DEM) assembly of virtual particles is calibrated
to approximate the behavior of a natural sand in undrained loading.
The particles are octahedral, bumpy clusters of spheres which are
compacted into assemblies of different densities.
The contact model is a J\"{a}ger generalization of the Hertz contact,
yielding a small-strain shear modulus that is proportional
to the square root of confining stress.  
Simulations of triaxial
extension and compression loading conditions and of simple shear 
produce behaviors 
that are similar to sand.
Undrained cyclic shearing simulations are performed with non-uniform
amplitudes of shearing pulses and with 24 irregular seismic shearing sequences.
A methodology is proposed for quantifying the severities of
such irregular shearing records, allowing the 24 sequences to be ranked
in severity.
The relative severities of the 24 seismic sequences
show an anomalous dependence on sampling density.
Four scalar measures are proposed for predicting the severity
of a particular loading sequence.
A stress-based scalar measure shows superior efficiency
in predicting initial liquefaction and pore pressure rise.
\end{abstract}
\KeyWords{Liquefaction, discrete element method, contact mechanics, 
          simulation, undrained loading.}
\section{Introduction}
Cyclic liquefaction is commonly thought to develop from the micro-scale
jostling of particles during repeated load reversals
or rotations of the principal stresses, causing a progressive
rearrangement of the particles and a tendency of the soil to contract.
This tendency, under undrained conditions, produces positive pore
pressure, which leads to a reduction in effective stress and a diminished
capacity of the particles to sustain load.
In the context of understanding soil behavior at the micro-scale,
rather than at the meta-scale of continuum constitutive approaches, the
micro-level
basis of liquefaction was confirmed in the particle-scale discrete
element (DEM) simulations of \citeN{Hakuno:1988a}
and \citeN{Dobry:1992a}. Over twenty years old, these
simulations of two-dimensional arrays of disks and spheres may seem
inelegant by current standards, but they gave convincing demonstration
of a micro-scale origin of cyclic loading behavior: pore pressure
rise concurrent with loading and a degradation of the shear modulus
with increasing strain magnitude. In a later series of two-dimensional
simulations, \citeN{Ashmawy:2003a} produced realistic
liquefaction curves, giving the relationship between cyclic stress
amplitude and the number of cycles to failure. Other simulations have shown
that the load-bearing capacity of a granular material is diminished
during repeated loading, reducing the number of inter-particle contacts,
leading to pore pressure rise under undrained conditions,
and altering the fabric anisotropy \cite{Ng:1994a,Sitharam:2003a,Sazzad:2010a}.
Recently, the interplay of pore fluid and grains has been simulated
by coupling DEM with discretized Navier-Stokes models, permitting the
simulation of entire soil strata to track the progression of larger-scale
phenomena such as lateral spreading and the upward migration of water
during ground shaking.
\par
The current work uses the Discrete Element Method (DEM) to explore
the cyclic liquefaction behavior of a target sand (Nevada Sand), attempting
a modest fidelity to its measured, laboratory behavior.
After showing that the model captures many aspects of this sand's
behavior, we use the model to simulate conditions that can occur in
the field but would be difficult to manage in a laboratory setting.
We arrive at a predictive measure of loading conditions
conducive to liquefaction.
\par
\begin{figure}
  \centering\includegraphics[width=3.3in]{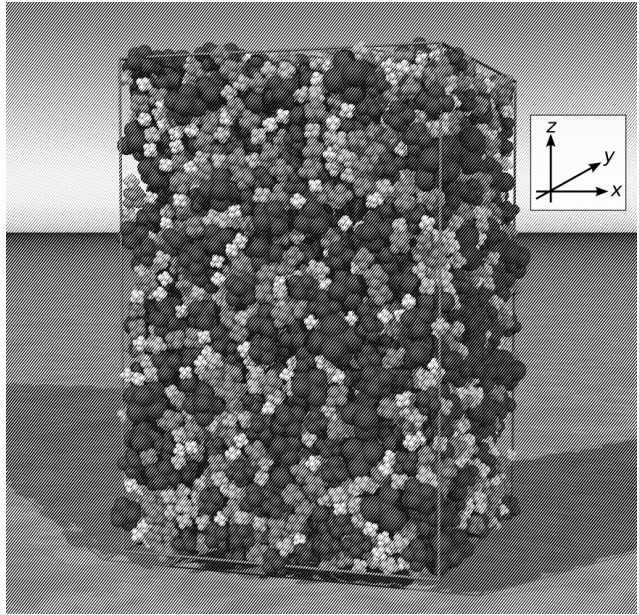}
  \caption{Discrete element (DEM) assembly of 6400 particles.%
         \label{fig:6400particles}}
\end{figure}

The DEM simulations in this study 
used the open source OVAL code \cite{Kuhn:2002b}
and are \emph{element tests}, in which
small assemblies of ``particles in a box'' undergo various deformation
sequences. The purpose is to explore the material behavior of a simulated
soil element~--- representing an idealized material point in a soil
continuum or an integration point in a
finite element model~--- rather than to study a larger
boundary value problem (e.g., a footing foundation or an entire soil
column, as with \citeNP{ElShamy:2012,ElShamy:2005a}). 
Figure~\ref{fig:6400particles}
shows an assembly of 6400 particles, representing a small soil element
of size $18\times12\times12\, D_{50}$ (about $3\times2\times2$mm):
large enough to capture the average material behavior but sufficiently
small to prevent meso-scale localization, such as shear bands, or
the macro-scale non-uniformities produced by boundary conditions (footings,
excavations, etc.).
In the study, we exploit certain advantages of DEM simulations for
exploring material behavior. 
Once a DEM assembly is created, the same
assembly can be reused with loading sequences of almost unlimited variety,
all beginning from precisely the same particle arrangement, during
which all Cartesian components of stress and strain are accessible.
DEM element tests also permit loading sequences with arbitrary control
of any six components of the stress and strain rates or of their linear
combinations---loading conditions that would require different testing
apparatus in a physical laboratory. 
The average stress within an assembly is computed from the inter-granular
contact forces, so that the computed stresses are inherently effective
stresses.
\par
The following section presents details of the DEM model, 
focusing on refinements
to current models. 
This section is followed by presentations of the model's monotonic and
cyclic loading behaviors. 
The cyclic response is explored for level-ground
conditions of cyclic simple shear in which shear stress is applied
in three types of sequences: uniform-amplitude loading, non-uniform amplitude
sequences, and realistic seismic loading sequences. 
These simulations are used to
evaluate proposed ``severity measures'' for predicting the onset of
liquefaction.
\section{Granular Assembly}
DEM assemblies 
(Fig.~\ref{fig:6400particles}) 
were constructed with the goal of
approximating the behavior of Nevada Sand, 
a standard poorly graded sand (SP)
used in laboratory and centrifuge testing programs, including
the VELACS program \cite{Arulanandan:1993a,Arulmoli1992a,Cho:2006a,Duku:2008a}.
A DEM model can be customized
by adjusting several characteristics, including (1)~particle size
and size distribution, (2)~particle shape, (3)~compaction procedure,
(4)~the contact force-displacement relation, and (5)~the contact friction
coefficient. 
At the outset, we recognized that a DEM model is unlikely
to reproduce all behaviors of a targeted soil, and we set modest goals
relative to Nevada Sand: similarities in particle size distribution,
range of void ratios, small-strain stiffness, and critical state friction
angle. 
\par
\begin{figure}
  \centering\includegraphics{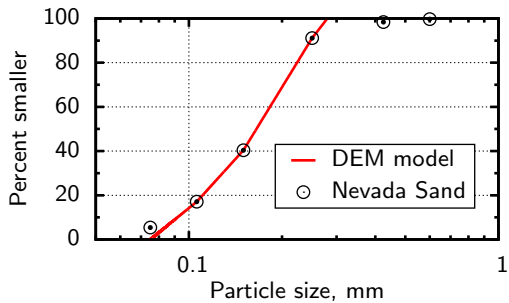}
  \caption{Particle size distributions.\label{fig:Particle-size}}
\end{figure}

Attaining the desired median particle size $D_{50}=0.165$mm
is a simple matter of scaling the DEM particles, but fashioning the
size distribution involves some compromise, since computation time
is favored by a smaller range of particle sizes. 
For this reason,
particle sizes were selected to fit the central portion of the particle
size distribution of Nevada Sand (Fig.~\ref{fig:Particle-size}),
neglecting the smallest and largest 3.5\% of sizes.
\par
\begin{figure}
  \centering\includegraphics[scale=0.2]{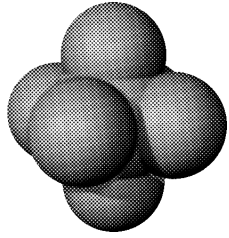}
  \caption{Particle shape: a bumpy octahedral cluster of spheres.
           \label{fig:bumpy}}
\end{figure}

An assembly of spheres can not adequately represent a natural sand:
sphere packings have a narrow range of void ratios (typically $e=0.55$
to 0.73 for glass ballotini, \citeNP{Zettler:2000a}); a sphere can
touch a neighboring sphere only at a single contact; and sphere assemblies
have relatively low strength (a friction angle $\phi$ of about 20$^{\circ}$,
\citeNP{Cho:2006a}).
%
%
To achieve more realistic simulations,
we chose a bumpy, compound cluster shape having a large central
sphere with six embedded ``satellite'' spheres in an octahedral arrangement
(Fig.~\ref{fig:bumpy}). 
Besides its computational advantages, the
shape has sufficient non-roundness to produce a large range of initial
densities, and the basic shape can be modified to attain a targeted
range of densities (i.e., one can modify the relative radii of the
single central sphere and the outer satellite spheres as well as the
relative protrusions of the outer spheres). 
As guidance, we used the
work of \citeN{Cho:2006a}, who developed correlations between
a sand's particle shape and its strength and density range.
\citeN{Salot:2009a} studied the effect of DEM particle shape
and contact friction on density and strength, and they developed a
procedure for calibrating a DEM assembly to approximate the behavior
of a targeted sand. 
With this guidance and considerable trial and
error, we arrived at a shape that produced a realistic strength and
range of void ratios, as described below. 
This shape has a ratio of
central-to-satellite sphere radii of 0.75, and the satellite 
spheres were centered
at octahedral points located 0.925 of the radius of the inner sphere
from its center (Fig.~\ref{fig:bumpy}).
\par
In a laboratory setting, sand can be conditioned, placed, and compacted
in various ways to produce a desired density and fabric. 
Because many
of these laboratory procedures can not yet be simulated, we used a
simpler computational procedure that produces assemblies with a similar
range of densities as Nevada Sand and with a modest fabric anisotropy,
as would be expected with a laboratory pluviation procedure. 
To start,
the 6400 particles were sparsely and randomly arranged within a spatial
cell surrounded by periodic boundaries. 
In the absence of gravity
and with a reduced inter-particle friction coefficient ($\mu=0.30$),
the assembly was anisotropically (uniaxially) compacted by slowly reducing
its height but with no lateral strain. 
The initially sparse arrangement
with zero stress would eventually ``seize'' when a loose, yet load-bearing,
fabric had formed. 
A series of fourteen progressively denser assemblies
were created by repeatedly assigning random velocities to particles
of the previous assembly (simulating a disturbed or vibrated state)
and then further reducing the assembly height until the newer specimen had
seized again. 
The fifteen specimens had void ratios in the range $e_{\text{loosest}}$
to $e_{\text{densest}}$ of 0.850 to 0.525, a range that is similar
to that of Nevada Sand obtained with standard ASTM procedures. 
Although the authors do not contend that virtual 
specimens with a range $e_{\text{loosest}}$
to $e_{\text{densest}}$ correspond to the range $e_{\text{max}}$
to $e_{\text{min}}$ attained with ASTM procedures, some auxiliary
evidence does support a similarity in the two ranges. 
Attempting to
simulate glass ballotini, the same DEM procedure was applied to create
assemblies of spherical particles having a narrow range of diameters.
The simulated compaction procedure results in assemblies with the
range $e_{\text{loosest}}$ to $e_{\text{densest}}$ of 0.750 to 0.549,
which compares favorably with ranges $e_{\text{max}}$ to $e_{\text{min}}$
that have been reported for ballotini prepared with the ASTM procedures,
about 0.73 to 0.58 \cite{Zettler:2000a}.
\par
Having created fifteen assemblies with this anisotropic compaction
scheme, the friction coefficient $\mu$ was raised to $0.60$ and
each assembly was isotropically consolidated to a mean effective
stress of 10kPa.
This step simulates the isotropic consolidation of a pluviated sample,
as in standard triaxial testing, and leaves the sample with a
small initial anisotropy (a Satake fabric anisotropy 
$F_{zz}/F_{xx}=1.08$).
Most results in the paper involve a further isotropic consolidation
to the higher stress of 80kPa, so that results can be compared with
the Nevada Sand tests of \citeN{Arulmoli1992a}. In short,
the preparation initially created assemblies with an anisotropic fabric
at low stress, followed by isotropic consolidation to a mean effective
stress of 10kPa or higher.
\par
The small-strain behavior of a DEM assembly is sensitive to the particular
force-displacement model of the contacts. 
During cyclic loading of
a sand, the mean effective stress $p$ can be progressively reduced to nearly
zero, and further cyclic loading causes $p$ to rise and fall across
a broad range of values. 
We believe that the proper simulation of liquefaction
requires a contact model that appropriately reflects a sand's small-strain
material behavior over a range of $p$. 
As a minimum, the relationship
between small-strain bulk shear modulus $G_{\text{max}}$ and mean effective
stress $p$ should comport with that of sand. 
The modulus $G_{\text{max}}$
of sands is usually found to vary in proportion to $p^{\beta}$, where
exponent $\beta$ is in the range 0.4--0.6, depending on particle
shape \cite{Cho:2006a}, particle size gradation \cite{Wichtmann:2009a},
surface roughness \cite{Santamarina:1998a}, and pre-load conditioning.
A $\beta$ of 0.5 is commonly used in geotechnical practice and for
correlations among $G_{\text{max}}$, $e$, and $p$ \cite{Hardin:1978a}.
A exponent of 0.5 also fits the data for Nevada Sand \cite{Arulmoli1992a}
and was the targeted exponent in our study.
\par
From a micro-mechanics viewpoint, exponent $\beta$ is known to depend
upon the contact stiffnesses of particle pairs 
\cite{Walton:1987a,Goddard:1990a,Agnolin:2007c}.
Most DEM simulations use a standard Hertz-Mindlin contact model in
which particles touch at spherical surfaces and behave as elastic
bodies. 
This contact model gives a normal force $f^{\text{n}}$ that
is proportional to the normal contact indentation $\zeta$ raised
to the power $3/2$: as $f^{\text{n}}\propto\zeta{}^{3/2}$. 
The bulk
stiffness of a granular assembly can be estimated from a simple idealization
in which all contacts bear an equal force and the particle-scale displacement
field conforms with the bulk field. 
This simple model predicts an
exponent $\beta=1/3$, such that $G_{\text{max}}\propto p^{1/3}$
\cite{Walton:1987a}. 
Simulations of sphere assemblies, in which these
simplifying assumptions are removed, yield somewhat greater exponents
$\beta$ \cite{Agnolin:2007c}, and our own DEM simulations of sphere
assemblies give the proportionality $G_{\text{max}}\propto p^{0.42}$
(Table~\ref{tab:Gmax}, row~2). 
Simulations with the ``bumpy'' clusters
of Figs.~\ref{fig:6400particles} and~\ref{fig:bumpy} 
give $G_{\text{max}}\propto p^{0.39}$
(Table~\ref{tab:Gmax}, row~3). 
In these simulations, the grains
were assigned a shear modulus $G_{\text{s}}$ of 29GPa and a Poisson
ratio $\nu_{\text{s}}$ of 0.15,
values that lie within the range of quartz \cite{Simmons:1965a,Mitchell:2005a}.
A friction coefficient $\mu = 0.60$, also within the range of quartz, was
chosen to fit the behavior of Nevada Sand \cite{Mitchell:2005a}.
Simulation values of bulk
stiffness $G_{\text{max}}$ were measured at shear strain $\gamma=0.001\%$.
These values are compared with those of
laboratory resonant column tests of Nevada Sand and correlations gained
from various sands, which give $G_{\text{max}}=71$--96MPa
(Table~\ref{tab:Gmax}, row~10 and footnotes~f and~g).
In short, simulations with the standard Hertz-Mindlin contact model yield
a poor match with the exponent $\beta$
and over-predict $G_{\text{max}}$ for
the range of pressures that typically apply in field liquefaction
situations.
\begin{table}
  \caption{Effect of the contact profile and particle shape on small-strain 
           bulk stiffness $G_{\text{max}}$ and exponent $\beta$ 
           (as $G_{\text{max}}\propto p^{\beta}$).\label{tab:Gmax}}
  \centering%
  \begin{tabular}{rcccccc}
  \hline 
   & Particle & Contact &  &  & $G_{\text{max}}$, MPa & Exponent\\
  Row & shape & contour & Source & $A_{\alpha}$ & at $p=80$kPa & $\beta$\\
  \hline 
  1 & spheres & spherical ($\alpha=2$) & theory$^{\text{a}}$ & $1/2R$ & 180$^{\text{a}}$ & 0.33\\
  2 & spheres$^{\text{b}}$ & spherical ($\alpha=2$) & DEM & $1/2R$ & 118 & 0.42\\
  3 & sphere clusters$^{\text{c}}$ & spherical ($\alpha=2$) & DEM & $1/2R$ & 170 & 0.39\\
  4 & spheres$^{\text{b}}$ & conical ($\alpha=1$) & theory$^{\text{d}}$ & 0.050$^{\text{d}}$ & 142$^{\text{d}}$ & 0.50\\
  5 & spheres & conical ($\alpha=1$) & DEM & 0.050 & 138 & 0.56\\
  6 & sphere clusters$^{\text{c}}$ & conical ($\alpha=1$) & DEM & 0.070 & 89.6 & 0.56\\
  7 & sphere clusters$^{\text{c}}$ & $\alpha=1.3$ & DEM & 5.3$\,^{\text{e}}$ & 90.2 & 0.50\\
  8 & sphere clusters$^{\text{c}}$ & $\alpha=0.8$ & DEM & $0.0045$$\,^{\text{e}}$ & 90.2 & 0.60\\
  9 & sphere clusters$^{\text{c}}$ & $\alpha=2.1$ & DEM & $5.5\times10^{5}$$\,^{\text{e}}$ & 89.6 & 0.40\\
  10 & sand & --- & experiment & --- & 71--96$\,^{\text{f}}$ & 0.4--0.6$\,^{\text{g}}$\\
  \hline 
  $^{\text{a}}$ & \multicolumn{6}{l}{\parbox[t]{6.0in}
    {See \citeN{Walton:1987a}. 
     An estimate of $G_{\text{max}}$ depends upon packing characteristics. 
     The values shown correspond to packing conditions of Row~2.} }\\
  $^{\text{b}}$ & \multicolumn{6}{l}
    {DEM assembly of 6400 spheres with $e=0.538$, $G_{\text{s}}=29\times10^{9}$GPa,
     and $\nu_{\text{s}}=0.15$.}\\
  $^{\text{c}}$ & \multicolumn{6}{l}
    {Figs.~\ref{fig:6400particles} and~\ref{fig:bumpy}. Assembly of
     6400 particles with $e=0.638$, $G_{\text{s}}=29\times10^{9}$GPa,
     and $\nu_{\text{s}}=0.15$.}\\
  $^{\text{d}}$ & \multicolumn{6}{l}{\parbox[t]{6.0in}{See \citeN{Goddard:1990a}. An estimate of $G_{\text{max}}$ depends upon packing characteristics.  The values shown correspond to the packing characteristics of the assembly in Row~2.}}\\
  $^{\text{e}}$ & \multicolumn{6}{l}{$A_{\alpha}$ chosen to yield $G_{\text{max}}\approx90$MPa. Values
    of $A_{\alpha}$ have dimensional units m$^{1-\alpha}$.}\\
  $\,^{\text{f}}$ & \multicolumn{6}{l}%
   {\parbox[t]{6.0in}{Resonant column testing of Nevada Sand 
    (specimen 60-43, $e=0.659$,
    $\gamma=0.001\%$, $G=71$MPa, \citeNP{Arulmoli1992a});
    correlations of \citeNP{Hardin:1978a}
    ($e=0.638$, $p=80$kPa, $G_{\text{max}}=88$MPa);
    and correlations of \citeNP{Wichtmann:2009a}
    ($e=0.638$, $C_{u}=2.1$, $p=80$kPa, $G_{\text{max}}=96$MPa).
    Also see \cite{Pestana:1995a}.}}\\
  $\,^{\text{g}}$ & \multicolumn{6}{l}{\parbox[t]{6.0in}{%
       See \citeNP{Cho:2006a,Wichtmann:2009a}. For Nevada Sand,
       $\beta=0.5$, \citeNP{Arulmoli1992a}, specimen 60-43.}}\\
  \hline 
  \end{tabular}
\end{table}
\par
\begin{figure}
  \centering\includegraphics{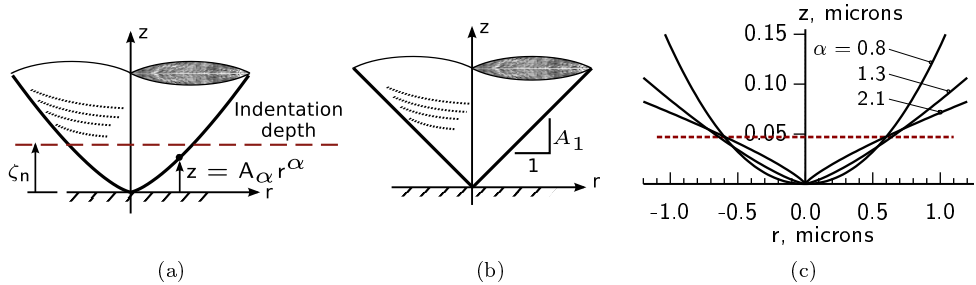}
  \caption{Contours of contact asperities as power-law surfaces of revolution:
           (a)~general power-law form with exponent $\alpha$ and parameter 
           $A_{\alpha}$, (b)~conical asperity, and (c)~asperity used 
           in the DEM simulations (labeled $\alpha=1.3$).\label{fig:Asperities}}
\end{figure}

\citeN{Goddard:1990a} has noted that a larger exponent $\beta$
is obtained if the particles interact at conical asperities rather
than along ideally smooth spherical surfaces (Fig.~\ref{fig:Asperities}b).
He arrived at an exponent $\beta=1/2$ (as $G_{\text{max}}\propto p^{1/2}$)
by applying the same simplifying assumptions that lead to a value
of $1/3$ for spherical contacts. 
Our DEM simulations with assemblies of spheres and of sphere
clusters having conical asperities give an exponent $\beta$ of 0.56
(Table~\ref{tab:Gmax}, rows~4 and~5), over-predicting our target
value of $\beta=0.50$ with both particle shapes. 
\par
Although the true nature of contact
between natural sand particles is currently a matter of intense interest
(see \citeNP{Cavarretta:2010a,Cole:2010a}), their contact surfaces
are certainly not glassy smooth spheres. 
We used a technique in which contacts were numerically detected at
the smooth spherical lobes of the bumpy clusters (Fig.~\ref{fig:bumpy});
whereas contact forces were computed by assuming rounded but
non-spherical asperities of about 1 micron width.
\citeN{Jager:1999a}
derived the normal force $f^{\text{n}}$ between an asperity of a
general form (i.e., a solid of revolution having the power-form
surface contour $z=A_{\alpha}r^{\alpha}$, for any positive $\alpha$)
and a hard flat surface (see Fig.~\ref{fig:Asperities}a):
\begin{equation}
  f^{\text{n}}=C_{\alpha}\zeta^{1+1/\alpha},\quad C_{\alpha}=\frac{4\alpha G_{\text{s}}}{(1-\nu_{\text{s}})(1+\alpha)}\left(\frac{\Gamma\left(\frac{1+\alpha}{2}\right)}{\sqrt{\pi}A_{\alpha}\Gamma\left(\frac{2+\alpha}{2}\right)}\right)^{1/\alpha}\label{eq:fn_jager}
\end{equation}
where $\zeta$ is the indentation depth (half of the contact overlap),
$G_{\text{s}}$ and $\mu_{s}$ are the shear modulus and Poisson ratio
of the solid grains, and $\Gamma$ is the gamma function. 
For smooth spherical surfaces that conform with a particle's radius
$R$, the exponent $\alpha$ is 2 and the contour parameter $A_{2}$
is $1/(2R)$, so that Eq.~(\ref{eq:fn_jager}) yields the standard
Hertz solution
\begin{equation}
  C_{2}=\frac{8}{3}\frac{G_{\text{s}}}{1-\nu_{\text{s}}}R^{1/2},\quad f^{\text{n}}=\frac{8}{3}\frac{G_{\text{s}}}{1-\nu_{\text{s}}}R^{1/2}\zeta^{3/2}
\end{equation}
With a conical asperity ($\alpha$=1, Fig.~\ref{fig:Asperities}b),
$A_{1}$ corresponds to the outer slope of the cone, and
\begin{equation}
  C_{1}=\frac{4G_{\text{s}}}{\pi(1-\nu_{\text{s}})}\frac{1}{A_{1}},\quad f^{\text{n}}=\frac{4G_{\text{s}}}{\pi(1-\nu_{\text{s}})}\frac{1}{A_{1}}\zeta^{2}\label{eq:jager_cone}
\end{equation}
By decoupling the asperity shape from the more general contour of
a particle's surface, Eqs.~(\ref{eq:fn_jager}) and~(\ref{eq:jager_cone})
afford a free parameter $A_{\alpha}$ that can be chosen so that the
DEM assembly has a $G_{\text{max}}$ similar to that of a targeted
sand. 
\par
To produce simulations in which exponent $\beta=0.50$ 
and $G_{\text{max}}\propto p^{0.50}$,
we used an asperity contour with parameter $\alpha=1.3$ (Eq.~\ref{eq:fn_jager}),
forming a rounded cone whose surface lies between spherical and conical
contours (Table~\ref{tab:Gmax} row~7 and Fig.~\ref{fig:Asperities}c).
The value 1.3 was chosen through trial and error, with the corresponding
parameter $A_{1.3}$ chosen so that $G_{\text{max}}$ is close to
the target value of 90MPa at a mean effective stress of 80kPa. The simulations
in the paper employ this pair of values, $\alpha$ and $A_{\alpha}$.
For particles of sub-mm size, such as in Nevada Sand, Eq.~(\ref{eq:fn_jager})
these conditions
imply an indentation depth of the asperities of a few tens of nano-meters
(about 0.05 microns for $p=80$kPa) and a width of about one micron
(Fig.~\ref{fig:Asperities}c). 
Alternative pairs of values $\alpha$
and $A_{\alpha}$, with shapes that are more rounded and more pointed
(Table~\ref{tab:Gmax}, rows~8 and~9, and Fig.~\ref{fig:Asperities}c)
yield the exponents $\beta=0.40$ and 0.60 respectively, encompassing
the range of small-strain behaviors that have been measured with 
sands (e.g. Table~\ref{tab:Gmax} footnote ``f''). 
\par
A DEM simulation must also compute the tangential forces between particles,
accounting both for elastic effects and for the frictional limit of
force. 
Although tangential contact motion is often idealized as advancing
steadily across a particle's surface, DEM simulations reveal that
tangential motions are quite irregular and errant and that the normal
force will irregularly increase and decrease during the concurrent tangential
motion \cite{Kuhn:2011a}. 
The calculation of tangential force between
DEM particles must account for the complex elastic-frictional response
during such irregular motions, particularly when an assembly is undergoing
realistic seismic loading. 
The tangential contact forces were calculated
with an extension of Hertz-Mindlin-Deresiewicz theory \cite{Mindlin:1953a}
by using the more general \emph{J\"{a}ger contact} algorithm 
\cite{Jager:2005a,Kuhn:2011a}.
This algorithm fully accounts for arbitrary sequences of normal and
tangential contact movements in a three-dimensional setting, while
maintaining the objectivity of the resulting contact forces. 
(The pseudo-code in \citeNP{Kuhn:2011a} requires the modification of only
two lines, 13 and 42, to accommodate the general Eq.~\ref{eq:fn_jager}.) 
\section{Monotonic Loading}
\begin{figure}
  \centering\includegraphics{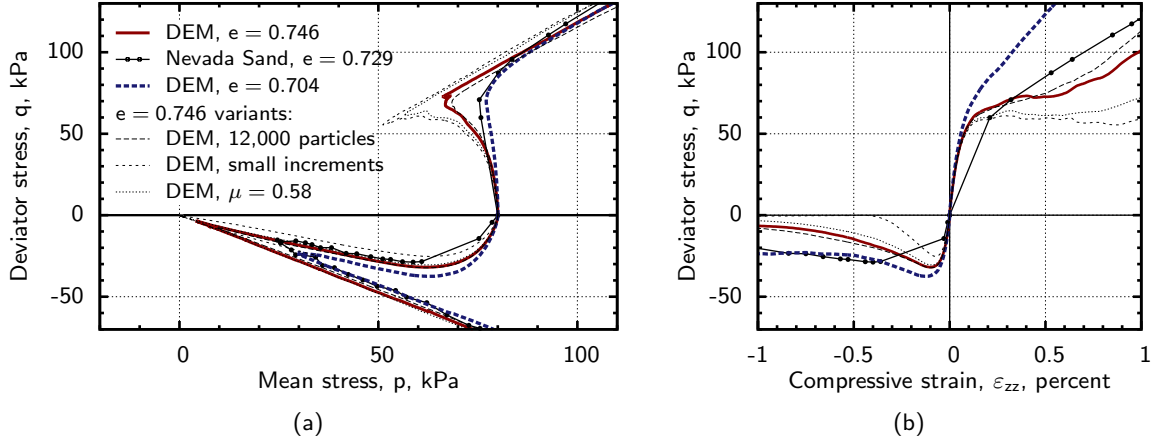}
  \caption{Undrained triaxial compression and extension of DEM simulations and
           Nevada Sand tests (tests CIUC40-04 and CIUE40-12 are shown,
           Arulmoli et al.\ 1992).}
           \label{fig:CIUC}
\end{figure}

Before conducting cyclic tests, the DEM model was calibrated and verified
by comparing its monotonic undrained loading behavior with that of
Nevada Sand. These tests were used to select the inter-particle friction
coefficient, $\mu=0.60$. Figure~\ref{fig:CIUC} shows stress paths
of undrained triaxial compression and extension simulations with assemblies
having void ratios 0.704 and 0.746 as well as a laboratory test of
Nevada Sand with void ratio 0.734 (the $D_{r}=40\%$ tests of 
\citeNP{Arulmoli1992a}).
Heavier lines in Fig.~\ref{fig:CIUC} are for simulations conducted under
the conditions described above and presented throughout most of the paper.
Thinner lines are for variations of these conditions discussed in
the next paragraph.
In typical undrained laboratory tests, the pore fluid is entrapped
within a saturated soil sample, preventing volume change during loading.
The DEM model contains no interstitial fluid; instead, undrained,
zero volume-change conditions are created by prescribing normal strains
in the three coordinate directions: 
$(1+\varepsilon_{xx})(1+\varepsilon_{yy})(1+\varepsilon_{zz})=1$
(Fig.~\ref{fig:6400particles}).
The DEM assembly was consolidated from the initial mean effective stress of
10kPa to a mean stress $p_{\text{o}}=80$kPa, and the subsequently
induced ``pore water pressure'' was computed from measured reductions
in the mean effective stress: $\Delta u=p_{\text{o}}-p$,
where $p$ is directly computed from the interparticle forces. 
\par
Loading was applied in the $z$-direction
in a slow, quasi-static manner, with movements
of the periodic boundaries 
much slower than the material's wave speed. 
As in many DEM simulations,
time was used as a surrogate parameter that advances deformation from
one integration step to another, with sufficient steps to allow particles
to adjust to the advancing deformation, thus economizing the computational
run-time.
A particle density much smaller than that of sand minerals was used in the 
simulations, a common approach in DEM analysis, which reduces the number
of time steps while maintaining nearly quasi-static conditions
(e.g. \citeNP{Thornton:1998a,OSullivan:2004c}).
Strain increments $\Delta\varepsilon_{zz}=\pm 5\times 10^{-7}$
in triaxial compression and extension were sufficiently small
to maintain an average force imbalance per particle of less than
$4\times 10^{-3}$ times the average contact force and an average
assembly kinetic energy less than $5\times 10^{-4}$ of the
internal elastic energy.
Although nearly quasi-static, the simulations were not
rate-independent.
Reducing the increment $\Delta\varepsilon_{zz}$ in half
softened the behavior (Fig.~\ref{fig:CIUC}, ``small increments'' lines).
The effect is similar to reducing the friction coefficient
$\mu$ to 0.58 (also in Fig.~\ref{fig:CIUC}).
Consistent conditions of strain increment and friction coefficient were
used throughout all of the simulations described below.
A larger assembly of 12,000 particles was also tested (Fig.~\ref{fig:CIUC}),
but the results are nearly the same as those of the smaller assembly,
which is sufficient for modeling undrained behavior.
\par
The two DEM simulations in Fig.~\ref{fig:CIUC} (heavy lines)
are for specimens that straddle the
density of the Nevada Sand specimen, and these simulations 
capture the primary features found in the laboratory tests:
strongly strain-softening behavior during triaxial extension
that is arrested by phase-transformation at a stress $p\approx25$kPa.
At larger strains (Fig.~\ref{fig:CIUC}a), 
the stress paths of the simulations converge to
roughly the same critical-state slopes --- in both extension and compression
--- as those of Nevada Sand.
Because of these similarities, the same DEM parameters were applied in
the remaining simulations.
\par
\begin{figure}
  \centering\includegraphics{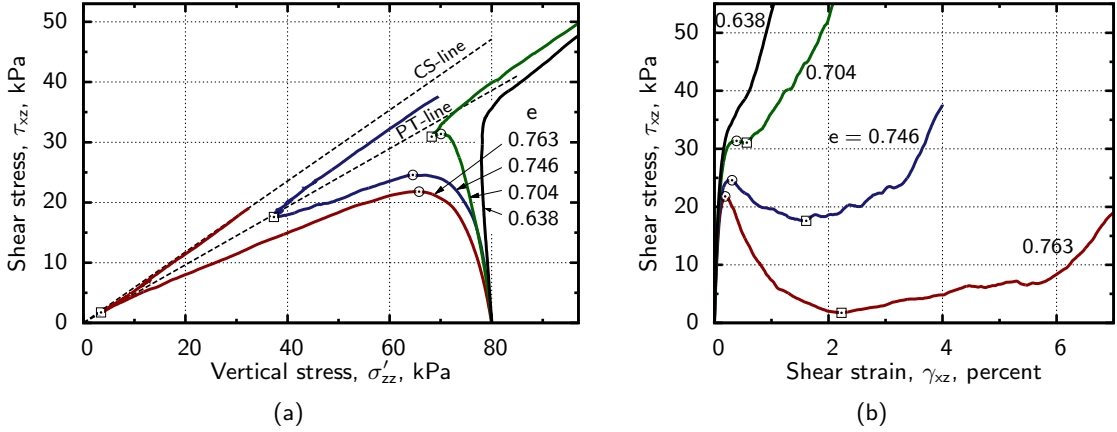}
  \caption{Undrained monotonic simple shear simulations,
           showing instability points ($\odot$), phase-transformation
           points ($\boxdot$), phase-transformation (PT) line,
           and critical-state (CS) line.\label{fig:SSU}}
\end{figure}

The undrained behavior in simple shear is illustrated in Fig.~\ref{fig:SSU}
for four DEM assemblies of different densities. 
These undrained tests started from an isotropic stress state,
and the shear strains were advanced monotonically with
$\dot{\gamma}_{xz}>0$ and 
$\varepsilon_{xx}=\varepsilon_{yy}=\varepsilon_{zz}=\gamma_{xy}=\gamma_{yz}=0$
(see Fig.~\ref{fig:6400particles}), as might be applied in
hollow-torsion constant-height undrained laboratory tests. 
Unlike the triaxial conditions of Fig.~\ref{fig:CIUC}, the directions
of the principal stresses rotated during the shear loading.
Markers locate the instability
points ($\odot$) at which shear stress $\tau_{xz}$ reached a temporary
peak and the phase-transformation points ($\boxdot$) at which the
vertical effective stress $\sigma^{\prime}_{zz}$ was minimum, 
a state commonly ascribed to a transition
from compressive to dilatant behavior. 
The two loosest assemblies
have stress paths that display temporary instability,
as would be
expected with loose dry-pluviated clean sands, and these looser assemblies
have more contractive behavior and lower instability and phase-transformation
points than those of the denser assemblies. 
An interpreted phase-transformation
line (PT-line) is shown in Fig.~\ref{fig:SSU}a, although the stress ratios
$\tau_{xz}/\sigma^{\prime}_{zz}$ 
of the four PT points decrease slightly with increasing
assembly density. 
\par
In a complementary series of \emph{drained} simple-shear
constant-$\sigma_{zz}$
simulations on the same assemblies,
we observed transformations from
compressive to dilatant behaviors 
for the three loosest assemblies~--- a transition
called the characteristic-state (ChS) \cite{Ibsen:1999a}. 
Our results
show that the same stress ratios $\tau_{xz}/\sigma^{\prime}_{zz}$ apply to both PT and
ChS transitions for the three assemblies, although the characteristic-state
occurs at larger shear strains. 
The same drained simple-shear constant-$\sigma_{zz}$
simulations were also used to evaluate the critical-state, a condition
that is attained at large strains and at which shearing progresses
at constant density and shear stress. 
The critical-state was reached
at shear strains $\gamma_{xz}$ greater than 80\%, and the corresponding
ratio $\tau_{xz}/\sigma^{\prime}_{zz}$ is shown as the CS-line in 
Fig.~\ref{fig:SSU}a.
The critical-state void ratio is 0.912, much larger (looser) than
the initial densities of the four assemblies, a result that is consistent
with the PT transition to dilatant behavior observed in the undrained
simulations.
\section{Cyclic Simple Shear}
Three types of cyclic loading sequences were simulated: (1)~uniform
amplitude cyclic shearing, (2)~alternating and modulated sequences
of small and large cyclic amplitudes, and (3)~realistic, erratic
sequences of seismic shearing. 
In all cases, cyclic shearing was uni-directional
and conducted as undrained simple-shear in the horizontal $x$-direction
(i.e., with shearing strains $|\gamma_{xz}|>0$ and $\gamma_{yz}=\gamma_{xy}=0$,
Fig.~\ref{fig:6400particles}). 
As with monotonic loading simulations,
the assemblies were consolidated to an isotropic stress of 80kPa, 
and undrained conditions
were imposed by preventing normal strains in the three coordinate
directions: $\varepsilon_{xx}=\varepsilon_{yy}=\varepsilon_{zz}=0$.
No ambient shear stress was imposed, corresponding to level-ground
conditions. 
Pore water pressure was computed from the measured reductions
in mean effective stress: $\Delta u=p_{\text{o}}-p$. 
The paper focuses
primarily on four assemblies having a range of void ratios $e=0.638$
to 0.763, corresponding to relative densities $D_{r}$ in a range
of about 70\% to 35\%.
\subsection{Uniform amplitude cyclic shearing}
\begin{figure}
\centering\includegraphics{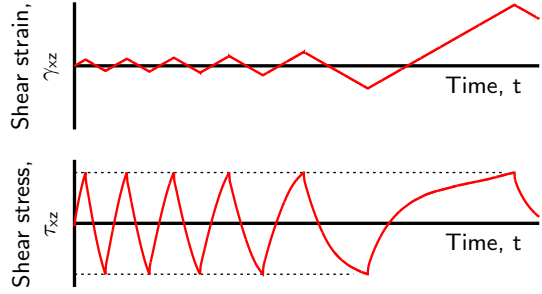}

\caption{Loading program for uniform-amplitude cyclic simple shear.\label{fig:zigzag}}
\end{figure}

\begin{figure}
  \centering\includegraphics{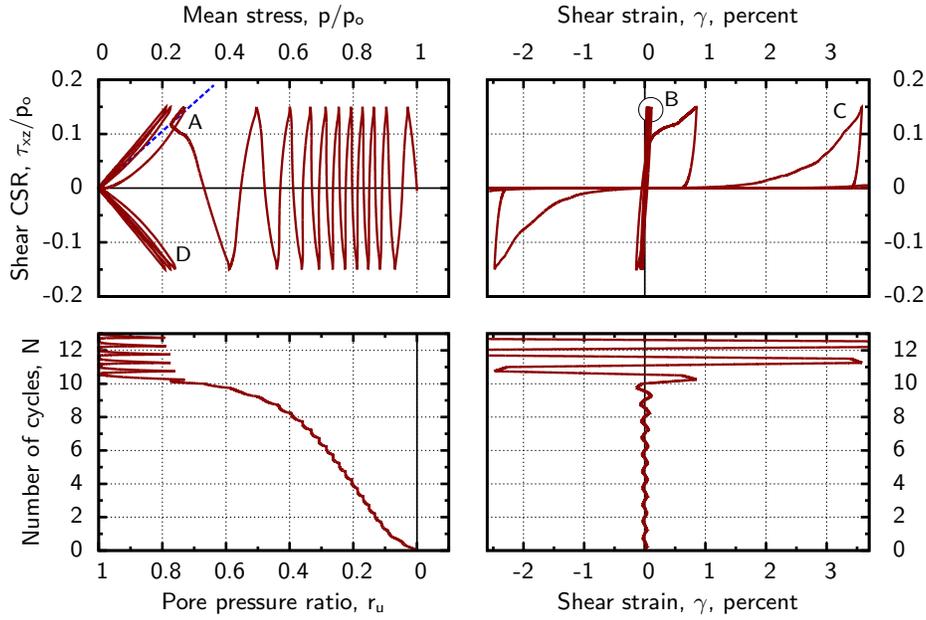}
  \caption{DEM simulation of cyclic undrained simple shear loading: $e=0.704$,
           $p_{\text{o}}=80$kPa, and $\tau_{xz}=13$kPa. The phase-transformation
           line is show in the stress path plot as a dashed line (blue in the
           electronic edition).\label{fig:4way_plot}}
\end{figure}

Undrained cyclic simple shear loading was applied in a sawtooth manner:
a uniform shearing rate $\pm\dot{\gamma}_{xz}$ was imposed in forward
and backward directions, reversing the direction each time a target
amplitude of shear stress $\tau_{xz}$ was reached (Fig.~\ref{fig:zigzag}).
 Loading proceeded until the mean effective stress had reached zero
--- at initial liquefaction --- and the total traversed strain $\int|d\gamma_{xz}|$
exceeded 10\%. 
These conditions repeatedly rotated and counter-rotated the principal
stress directions.
The mean stress and shear stress were recorded throughout
these strain-controlled histories. The four-way plot in Fig.~\ref{fig:4way_plot}
shows typical results, in this case with a cyclic stress amplitude
$\tau=\pm13$kPa 
(i.e., a cyclic stress ratio, CSR, $\tau_{xz}/p_{\text{o}}=0.163$).
These plots show the stress path, the stress-strain evolution, and
the pore pressure ratio $r_{u}=-\Delta u/p_{\text{o}}$, all of which
resemble those of saturated sands. 
The pore pressure increases steadily,
and at about 10~cycles, the stress path expresses
phase-transformation behavior (labeled ``A''), whereupon the mean
effective stress collapses to nearly zero. 
Once phase-transformation has occurred,
the stress-strain evolution changes from the narrow hysteresis pattern
of the first 9 cycles (labeled ``B'') into a broader scythe-shaped
pattern (labeled ``C'').  
Initial liquefaction ($r_{u}=1.0$) occurs after 10\textonehalf{}
cycles of loading; and a shear strain of 3\% is reached at 11~cycles. 
After liquefaction is initiated, the stress path falls into
butterfly repetitions, typical of sands (labeled ``D''). 
These results are qualitatively consistent with undrained
cyclic shear tests of sands
\cite{Arulmoli1992a,Kammerer:2000a,Porcino:2007a}.
\par
\begin{figure}
  \centering\includegraphics{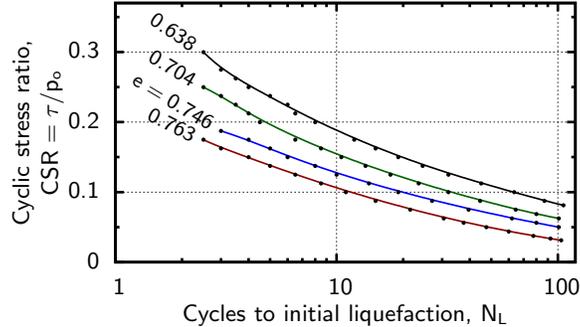}
  \caption{Liquefaction curves of four DEM assemblies.
           \label{fig:Liquefaction-curves}}
\end{figure}

Figure~\ref{fig:Liquefaction-curves} shows the liquefaction curves
obtained from multiple simulations of four assemblies having different
void ratios. The upward curvature in this semi-log plot is similar
to that of sands, although the curves have a steeper downward slope
than with most sands (e.g. \citeNP{Porcino:2007a}).
Confirming the choice of a rounded cone asperity profile
(Table~\ref{tab:Gmax}, Row~7),
simulations with a standard Hertz-Mindlin spherical contact
(Table~\ref{tab:Gmax}, Row~3) yielded an even steeper downward slope:
20--50\% more cycles at large shear stress ratios and 20--30\% fewer
cycles at small ratios.
We also obtained results for a large assembly of 12,000 particles, and
the results are nearly indistinguishable from those in 
Fig.~\ref{fig:Liquefaction-curves}.
\subsection{Non-uniform cyclic sequences}
\begin{figure}
  \centering\includegraphics{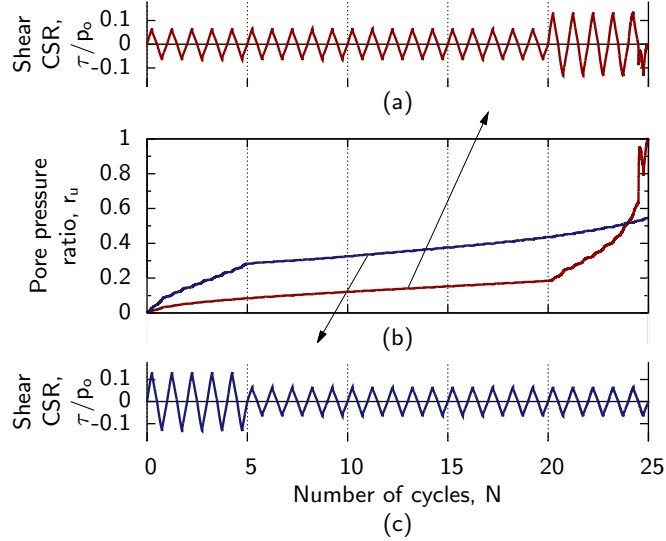}
  \caption{Simulations of two sequences of large and small cyclic 
           pulses ($e=0.746$, $p_{\text{o}}=80$kPa).
           \label{fig:bi-amplitude}}
\end{figure}

\citeN{Wang:1989a} conducted experiments on Monterey
\#1 Sand in which the amplitudes of cyclic pulses were either
increased or reduced during undrained loading. They found that the
final pore pressure depends on the sequencing of the variable-amplitude
cyclic pulses. We investigated this phenomenon with two types of simulations.
With the first type, two sequences of twenty-five pulses were applied:
five large-amplitude pulses were either 
preceded or followed by twenty small-amplitude
pulses having half the amplitude of the larger pulses (Fig.~\ref{fig:bi-amplitude}).
For sequences with magnitudes large enough to produce significant
pore pressures, we found that the more damaging sequences
start with the smaller pulses. Figure~\ref{fig:bi-amplitude} shows
typical results for a loose assembly with void ratio $e=0.746$. 
Through trial and error, we varied the reference amplitude (i.e., that of
the larger pulses), so that the full set of twenty-five 
pulses~--- twenty small followed by five large~--- would
produce initial liquefaction ($r_{u}=1$), maintaining the ratio 1:2
of pulse amplitudes (values $\tau/p_{\text{o}}=0.067$ and 0.134 in
the figure). 
Once the proper reference amplitude was established,
we ran the alternative sequence, with the larger pulses applied first.
This second sequence results in an $r_{u}$ of only 0.546. 
These observations are consistent with trends described by \citeN{Wang:1989a}. 
We note, however, that the difference in the 
effects of the two sequences is reduced
with denser assemblies, 
and a difference is almost non-existent with the densest
assembly ($e=0.638$).
\par
\begin{figure}
  \centering\includegraphics{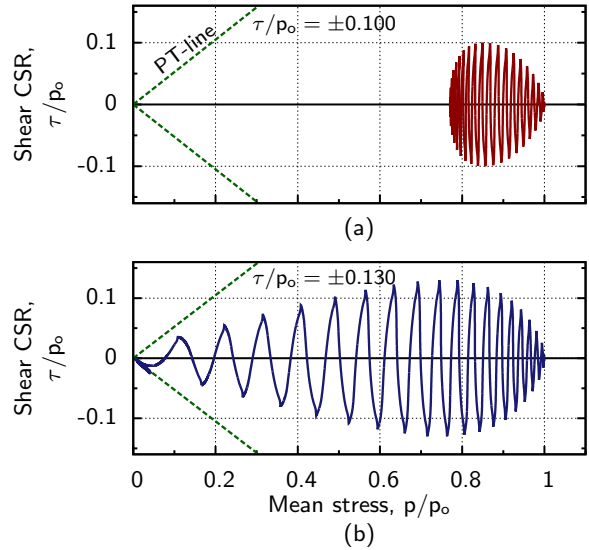}
  \caption{Simulations with modulated sequences of cyclic pulses ($e=0.746$,
           $p_{\text{o}}=80$kPa).\label{fig:modulated}}
\end{figure}

In a second type of simulation, we applied a modulated sequence of
rising and falling stress amplitudes (Fig.~\ref{fig:modulated}).
As with all other simulations, these were strain-controlled tests
in the manner of Fig.~\ref{fig:zigzag}, in which shear strain was
advanced at a constant rate $\pm\dot{\gamma}_{xz}$ until a target
shear stress was reached, whereupon the strain direction was reversed.
The target stress of the $i$th pulse was $\tau_{i}=\tau_{\text{max}}\sin(\pi i/N)$
for the $N=20$ modulated pulses, with 10 leading (rising) pulses
followed by 10 trailing (falling) pulses. When the maximum stress
$\tau_{\text{max}}$ is relatively small~--- producing a final $r_{u}$
less than 0.5~--- the leading pulses are more damaging than the trailing
pulses. This result is apparent in the stress path at the top 
of Fig.~\ref{fig:modulated},
where the stress path is more elongated to the right. With a larger
$\tau_{\text{max}}$, the opposite trend is observed: the trailing
pulses produce a larger rise in pore pressure, resulting in a stress
path that is elongated toward the left (bottom of Fig.~\ref{fig:modulated}).
\subsection{Seismic loading}
In a final series of simulations, we applied 24 transient seismic loading
sequences to four assemblies of 6400 particles having different void
ratios. 
By analyzing the simulation results, we propose a Severity Measure (SM)
that predicts the onset of liquefaction, based on shear stress records.
A suite of 24 ground motions was selected from the NGA database
maintained by the Pacific Earthquake Engineering Research (PEER) Center 
\cite{PEER:2000a}.
The selected ground motions were screened from about 4000 candidate
motions to provide a diversity of spectral and temporal conditions
as determined with four intensity measures (IMs): PGA/MSF (peak ground
acceleration with a magnitude scaling factor, e.g. \citeNP{Arango:1996a});
Arias intensity \cite{Kayen:1999a}; CAV$_{5}$ intensity (cumulative
absolute velocity, \citeNP{Kramer:2006a}); and NED intensity (normalized
energy demand, \citeNP{Green:2001a}). 
Each of the 24 motions produced
a large value of one IM but a low value of another IM, all in various
combinations of IM pairs, thus providing a suite of 24 motions with significantly
different amplitudes, frequency contents, durations, and phasing relationships.
\par
These ground acceleration records can not be input directly into the
DEM model. 
Shear stress histories were extracted from the ground accelerations
by applying these motions as inputs in an equivalent-linear wave propagation
model of a 6m sand layer using the ProShake software. 
The resulting
stress histories were in the form of cyclic shear stress ratio records 
(CSR records)
of shearing stress divided by the initial vertical confining stress,
$\tau_{xz}/\sigma_{zz,\text{o}}$ ($\tau_{xz}/p_{\text{o}}$ in
our isotropically consolidated simulations). 
Rather than applying
a CSR record directly, the record was processed in two ways. 
First, the CSR record was digitally perused to identify all of its reversals
of loading direction. 
These peaks and valleys became the target shearing
stresses at which the direction of the shearing strain $\pm\dot{\gamma}_{xz}$
was reversed while shearing with the same rate magnitude 
(see Fig.~\ref{fig:zigzag}).
\par
\begin{figure}
  \centering\includegraphics{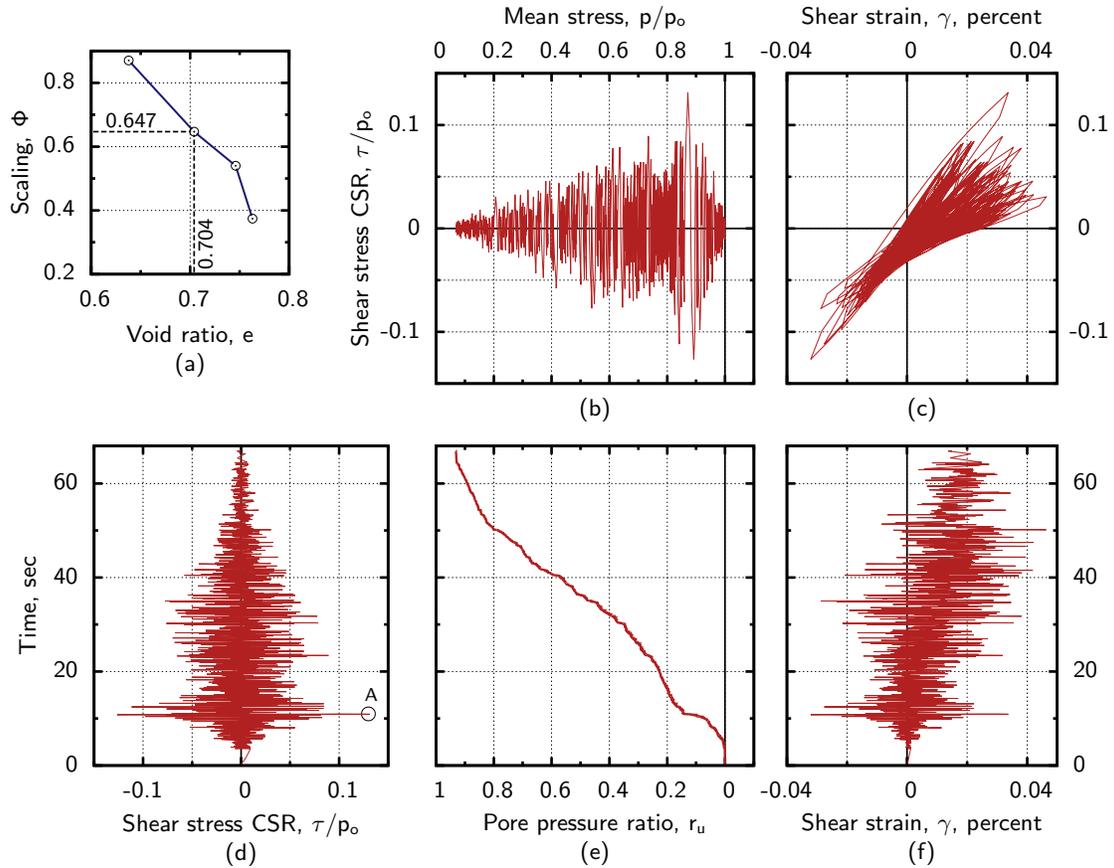}
  \caption{Seismic loading of a DEM assembly, $e=0.704$, $p_{\text{o}}=80$kPa.
           The record of cyclic stress ratios 
           ($\text{CSR}=\tau_{xz}/p_{\text{o}}$)
           is from the PEER NGA record \texttt{LANDERS/MCF000}. Shear stresses
           are scaled by factor $\Phi=0.647$ to suspend initial liquefaction
           until the end of the CSR record. The upper-left insert shows the 
           dependence
           of $\Phi$ on the assembly void ratio $e$.\label{fig:5way}}
\end{figure}

A second modification was applied at the start of a simulation: the
stresses of each CSR record were scaled by a factor $\Phi$ so that
initial liquefaction was delayed until the very end of the record.
A different scaling factor $\Phi$ was required for each of the 24
CSR records, and the factors would also differ among the four assemblies
having different void ratios. The necessary factors were determined
through a trial and error procedure for each of the 24 records and
for each void ratio. 
Figure~\ref{fig:5way} shows the results of
a single scaled CSR record for a DEM assembly with void ratio $e=0.704$.
The scaled record of CSR versus time is shown in the lower left of the plot.
The factor $\Phi=0.647$ in this figure causes the assembly to reach
an $r_{u}=0.932$ at the end of the record. 
Increasing $\Phi$ to 0.648 pushes the assembly
beyond initial liquefaction, producing a few small ``butterfly'' oscillations
at the end of the stress path (as in Fig.~\ref{fig:4way_plot}).
\par
Although arriving at the proper factors $\Phi$ is a time-consuming
process, this process serves three purposes:
\begin{enumerate}
\item
The factor $\Phi$ provides a quantifiable basis for ranking the severities
of the 24 original (unscaled) CSR records with respect to their propensity
for producing initial liquefaction ($r_{u}=1$). 
Specifically, the inverse
of each factor, $1/\Phi$, is a measure of the severity of the particular
ground motion and its shearing record (i.e., the original unscaled
CSR record). The 24 records are ranked in Fig.~\ref{fig:Ranking},
with the most severe records at the top and the most benign at the
bottom. The ranking in this figure was derived from the single assembly
with void ratio $e=704$. 
The scaled CSR record of Fig.~\ref{fig:5way}
appears near the middle of the ranking 
(labeled $\odot$ in Fig.~\ref{fig:Ranking}).
\begin{figure}
  \centering\includegraphics{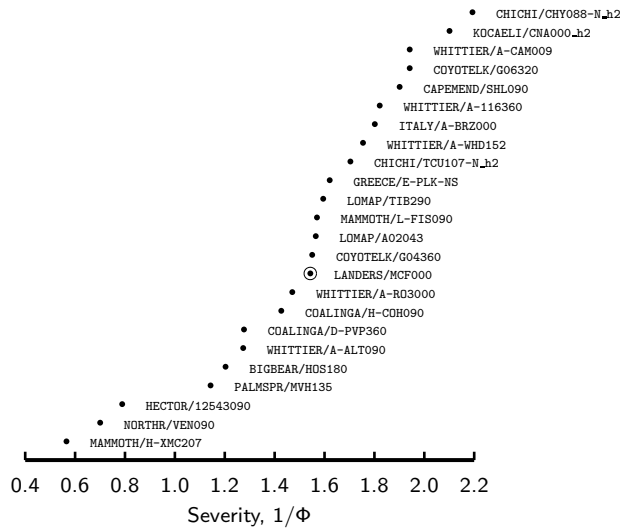}
  \caption{Ranking of severities $1/\Phi$ of 24 seismic CSR records ($e=0.704$,
           $p_{\text{o}}=80$kPa). Records near the top are the most severe,
           requiring a small $\Phi$ to forestall liquefaction until the end
           of the record.\label{fig:Ranking}}
\end{figure}
\item
Having scaled all 24 shear stress records so that each postponed
initial liquefaction until the end of the record, we can seek a commonality
in their (scaled) features. 
That is, we can explore possible severity
measures (SMs), which we define as a scalar value, derived from a
cyclic stress history (CSR record), that measures the record's propensity
for producing initial liquefaction. 
For example, the peak shear stress
in a CSR record (e.g., the single point ``A'' in the lower left
of Fig.~\ref{fig:5way}) could serve as a simple (albeit inefficient)
severity measure. 
An ideal SM would have the same scalar value for each
of the 24 scaled CSR records, since each record is scaled to reach
a common state of initial liquefaction at the end of the record. 
The range of the 24 SM values
for the scaled records is an indicator of the \emph{efficiency} of
a candidate SM. 
Although many candidate SMs were investigated, the
paper gives results for four SMs, described below.
%

\item
Besides its use as a predictor of initial liquefaction, an ideal severity
measure would also predict other damaging effects of a particular
seismic record. These effects could include pore pressure rise ($\Delta u$
or $r_{u}$) for CSR records that are not sufficiently severe to initiate
liquefaction, post-liquefaction strains for more severe records, etc.
\end{enumerate}
We address the first and second items and also apply a particular
Severity Measure (SM) to the prediction of pore pressure rise, as
suggested in the third item.
\par
The key to this approach is finding the scaling factor $\Phi$ of
each seismic CSR record that would postpone the onset of liquefaction
until the very end of the record. For this purpose, we made use of a primary
advantage of DEM simulations: the ability to repeatedly subject the
same assembly (i.e. virtual specimen) to the 24 records, each with
different scaling factors $\Phi$, thus finding the proper factors by
trial and error. 
Ten or eleven trials were usually necessary with
each CSR record to find its $\Phi$ with a precision of $\pm0.001$.
This same procedure was applied to all four assemblies having different
void ratios.
\par
\begin{figure}
  \centering\includegraphics{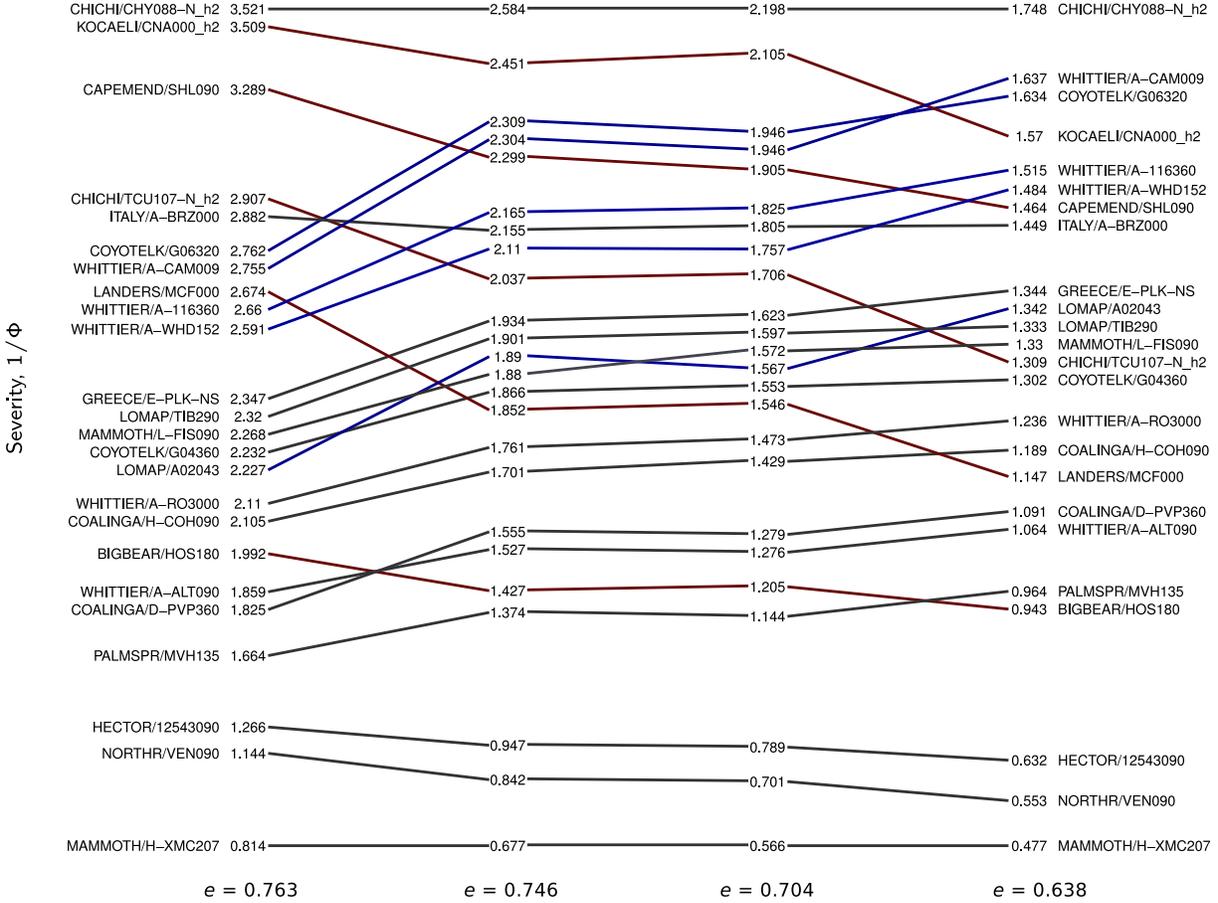}
  \caption{Slope-graph showing the rankings of severities $1/\Phi$ of 24 seismic
           CSR records, based upon simulations with four densities.
           \label{fig:Slope-graph-10-11-13-15}}
\end{figure}

The severities of the 24 records are shown in the slope-graph of 
Fig.~\ref{fig:Slope-graph-10-11-13-15}
for the four assemblies. 
In this figure, the records are ranked from the most severe (top,
large $1/\Phi$) to the least severe (bottom), with density increasing
from left to right
(the ranking in Fig.~\ref{fig:Ranking} is reproduced as the third
column in Fig.~\ref{fig:Slope-graph-10-11-13-15}). 
Because denser assemblies are more resistant to
initial liquefaction, the scaling factor $\Phi$ of each CSR record
must be increased with each increase in density (for example, with
the \texttt{CHICHI/CHY088-N\_h2} record at the top of 
Fig.~\ref{fig:Slope-graph-10-11-13-15},
the inverse factor $1/\Phi$ is reduced from 3.521 to 1.748 as the
void ratio decreases from 0.763 to 0.638). 
The ranking of the 24 records
is not consistent across the four densities, as is apparent from the
crossing lines. 
Oddly, the severities of certain CSR records, relative
to other records, decrease with increasing density (these records
are marked red in the electronic version); whereas, the severities
of other records increase relative to other records at greater density
(marked blue). 
For example, the \texttt{LANDERS/MCF000} record is
much more severe than the \texttt{LOMAP/A02043}
record when applied to the loosest assembly; but these roles are reversed
with the densest assembly. 
This anomalous density-dependent behavior
was also noted in a previous section regarding non-uniform sequences of large
and small shearing pulses.
\par
Many scalar severity measures (SMs) were explored as candidates for
predicting the propensity of a particular cyclic stress ratio (CSR) record
for producing initial liquefaction.
Four representative SMs are as follows:
\begin{align}
\text{SM}_{1}= & \left|\tau_{\text{max}}/p_{\text{o}}\right|\label{eq:SM1}\\
\text{SM}_{2}= & \int\frac{\tau\, d\gamma^{\text{plastic}}}{p_{\text{o}}}\label{eq:SM2}\\
\text{SM}_{3}= & \int H(|\gamma|-\gamma_{\text{t}})\,\left|d\gamma\right|\label{eq:SM3}\\
\text{SM}_{4}= & \int\left|d\left[ \left(\frac{|\tau|}{p}\right)^{2}\right]\right|
 = 2\int\frac{|\tau |}{p}\,\left| \frac{d|\tau |}{p}-\frac{|\tau |}{p}\frac{dp}{p}\right|\label{eq:SM4}
\end{align}
which represent the maximum shear ratio of a CSR record (SM$_{1}$),
a normalized energy demand (SM$_{2}$), a strain-path measure (SM$_{3}$),
and a stress-path measure (SM$_{4}$). For the uni-directional loading
of our DEM simulations, $\tau$ is the shear stress $\tau_{xz}$;
$\gamma$ is the shear strain $\gamma_{xz}$; $\gamma_{\text{t}}$
is a threshold shear stress (assumed to be 0.01\%); $H()$ is the
Heaviside function, which equals zero unless the current strain magnitude
$|\gamma_{xz}|$ exceeds $\gamma_{\text{t}}$ (in which case, $H=1)$;
$p_{\text{o}}$ is the initial mean effective stress; and $p$ is the current
mean effective stress. 
The plastic strain increment $d\gamma^{\text{plastic}}$
in Eq.~(\ref{eq:SM2}) is computed by subtracting the elastic increment
$d\tau/G_{\text{max}}$ from the full strain increment $d\gamma$,
where modulus $G_{\text{max}}$ is estimated with the relation given
in Table~\ref{tab:Gmax}, row~7. 
Unlike earthquake \emph{intensity} measures
such as the Arias intensity, these four SMs are not based upon ground motions
(accelerations or velocities) but instead are integrals of the stresses
and strains that result from these ground motions. 
We also note that the four SMs are rate-independent,
as time is not explicitly part of their definitions. 
The liquefaction resistance of sands is known to be insensitive 
to the loading rate
(i.e. nearly independent of excitation frequency), 
which is consistent with the four SMs. 
If a time history of shear stress or strain is available, the differential
quantities in Eqs.~(\ref{eq:SM2})--(\ref{eq:SM4}) can be replaced
with the corresponding rate differentials: for example $d\tau=(d\tau/dt)\, dt$.
\par
\begin{figure}
  \centering\includegraphics{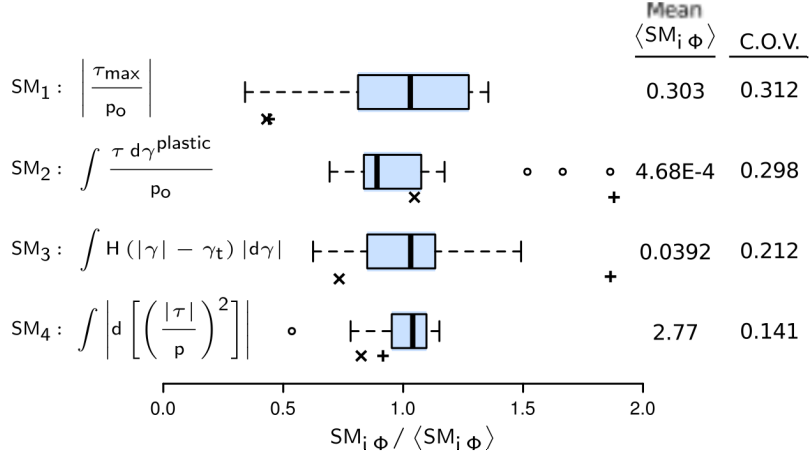}
  \caption{Efficiencies of four severity measures (SMs): box plots of 24 cyclic
           stress records, scaled to produce initial liquefaction ($e=0.746$,
           $p_{\text{o}}=80$kPa).  Values for the non-uniform sequence of
           Fig.~\ref{fig:bi-amplitude} (top) are marked ``$+$''; 
           values for the sequence
           of Fig.~\ref{fig:modulated} (bottom) are marked ``$\times$''.
           \label{fig:Efficiencies}}
\end{figure}

These four SMs were evaluated for the 24 seismic stress (CSR) records.
Each simulation yields a record of shear strain $\gamma_{xz}$ and
mean effective stress $p$ as well as the input stresses $\tau_{xz}$, permitting
evaluation of integrals (\ref{eq:SM2})--(\ref{eq:SM4}). 
As stated before, a scaling factor $\Phi$ was determined for each CSR record
that would delay initial liquefaction until the end of the record,
and the resulting 24 \emph{scaled} SM values correspond to a common
state of initial liquefaction ($r_{u}=1$), as will be denoted with a subscript
``$\Phi$''. 
Figure~\ref{fig:Efficiencies} shows box plots of the
four SMs in which their values from the 24 scaled CSRs ($\text{SM}_{i\Phi}$,
$i=1,2,3,4$) are normalized by dividing by the mean of the particular
SM, denoted as $\left\langle \text{SM}_{i\Phi}\right\rangle$.
The scatter in the simplest measure, $\text{SM}_{1\Phi}$,
is considerable, indicating that maximum shear stress is a poor
predictor of liquefaction.
Although $\text{SM}_{2\Phi}$, $\text{SM}_{3\Phi}$, and $\text{SM}_{4\Phi}$
exhibit smaller dispersions, the stress-path measure $\text{SM}_{4}$
has the least scatter, indicating a superior efficiency in predicting
initial liquefaction. 
The efficiencies of the four SMs are
summarized in the inset of Fig.~\ref{fig:Efficiencies}, giving
their coefficients of variation (standard deviation divided by mean),
with smaller coefficients corresponding to a more efficient and less
scattered severity measure. 
The measure $\text{SM}_{4}$ of Eq.~(\ref{eq:SM4})
yields the lowest dispersion and serves as an efficient predictor
of initial liquefaction.
\par
Figure~\ref{fig:Efficiencies} also shows results of applying
the four severity measures to the non-uniform cyclic sequences illustrated
in Figs.~\ref{fig:bi-amplitude} and~\ref{fig:modulated}.
Two of these sequences resulted in liquefaction (Fig.~\ref{fig:bi-amplitude}
top and Fig.~\ref{fig:modulated} bottom).
Although three of the severity measures were poor predictors of
liquefaction (the ``$+$'' and ``$\times$'' 
symbols in Fig.~\ref{fig:bi-amplitude}), 
the fourth measure $\text{SM}_{4}$ gave values close
to the threshold liquefaction value
$\left\langle \text{SM}_{4\,\Phi}\right\rangle$.
In contrast, the two-amplitude sequence at the bottom 
of Fig.~\ref{fig:bi-amplitude} did not result in liquefaction, a result
that is consistent with its low SM value:
$\text{SM}_{4}/\left\langle \text{SM}_{4\,\Phi}\right\rangle=0.54$.
\par
\begin{figure}
  \centering\includegraphics{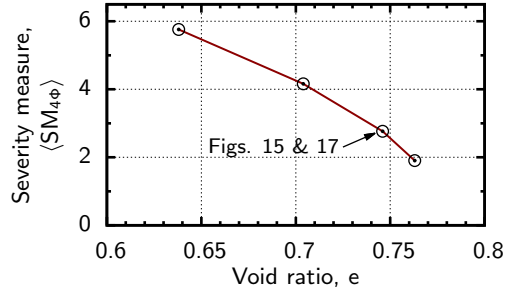}
  \caption{Effect of specimen density on the average value 
           of severity measure $\langle\text{SM}_{4\Phi}\rangle$
           (Eq.~\ref{eq:SM4}) at initial liquefaction.\label{fig:SM4_vs_e}}
\end{figure}

The value of an SM required to initiate liquefaction 
will depend upon a soil's density.
Figure~\ref{fig:SM4_vs_e} gives the 
values of SM$_{4\Phi}$ for four specimens having different void ratios,
based upon the averaged results of the 24 seismic sequences. 
As would be expected, the value of SM$_{4}$ required to initiate
liquefaction (i.e., SM$_{4\Phi}$) increases with increasing specimen density.
\par
\begin{figure}
  \centering\includegraphics{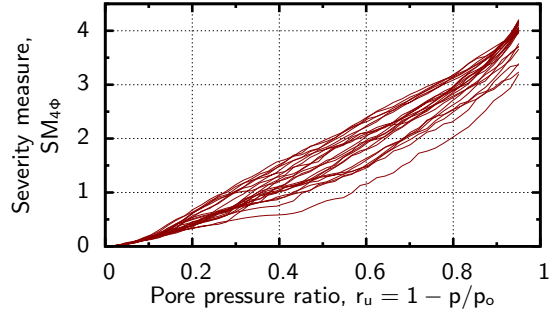}
  \caption{Relationship between severity measure SM$_{4\Phi}$ (Eq.~\ref{eq:SM4})
           and excess pore pressure ratio $r_{u}$ 
           ($e=0.746$, $p_{\text{o}}=80$kPa).\label{fig:SM4_vs_ru}}
\end{figure}

A proper measure of the severity of a cyclic sequence should also
predict the pre-liquefaction rise in pore pressure. 
Figure~\ref{fig:SM4_vs_ru}
shows the relationship between the excess pore pressure ratio 
$r_{u}=1-p/p_{\text{o}}$
and SM$_{4}$ for a single assembly subjected to the 24 seismic records.
This severity measure is a monotonically increasing function of
the shear stress history (scaled CSR record), which is
seen to advance in a roughly linear manner with increasing pore pressure.
Figure~\ref{fig:SM4_vs_ru} shows only modest scatter in the SM$_{4}$
vs. $r_{u}$ behavior, indicating that this severity measure would
serve as an efficient predictor of pore pressure rise.
\section{Concluding Remarks}
A discrete element (DEM) assembly of virtual particles has been calibrated
to approximate the behavior of a natural sand, particularly at small strains.
The paper presents simulation methodologies for exploring the complex
response of such granular materials to undrained cyclic loading.
Methods are also proposed for using simulations to rank 
the severities of different
seismic sequences and for developing scalar predictors of the severity.
Some anomalous behaviors have been observed, and a promising scalar
predictor of liquefaction susceptibility is identified.
Even though laboratory tests are the final arbiter of a
material's behavior, DEM simulations offer certain capabilities
that are difficult to achieve in a laboratory setting: in particular,
the ability to subject the same virtual assembly to nearly
unlimited loading sequences.
\par
Natural extensions of the current work would include simulations of
bi-directional seismic shearing and of seismic loading in
sloping-ground conditions.
Even with its advantages, DEM simulations continue to be hampered
by the computational demands of effectively simulating large, realistic
boundary-value problems (foundations, excavations, etc.) or
even conducting small element tests well into the post-liquefaction
regime in which the strain excursions become very large.
We believe, however, that discrete element 
simulations can serve to investigate many important aspects of
the complex cyclic behavior of soils.
\section*{Acknowledgement}
This material is based upon work supported by the National Science
Foundation under Grant No. NEESR-936408.
%

%
\end{document}